# Fast Latent Factor Analysis via a Fuzzy PID-Incorporated Stochastic Gradient Descent Algorithm

Jinli Li, Ye Yuan, *Member, IEEE*

*Abstract*—A high-dimensional and incomplete (HDI) matrix can describe the complex interactions among numerous nodes in various big data-related applications. A stochastic gradient descent (SGD)-based latent factor analysis (LFA) model is remarkably effective in extracting valuable information from an HDI matrix. However, such a model commonly encounters the problem of slow convergence because a standard SGD algorithm learns a latent factor relying on the stochastic gradient of current instance error only without considering past update information. To address this critical issue, this paper innovatively proposes a Fuzzy PID-incorporated SGD (FPS) algorithm with two-fold ideas: 1) rebuilding the instance learning error by considering the past update information in an efficient way following the principle of PID, and 2) implementing hyper-parameters and gain parameters adaptation following the fuzzy rules. With it, an FPS-incorporated LFA model is further achieved for fast processing an HDI matrix. Empirical studies on six HDI datasets demonstrate that the proposed FPS-incorporated LFA model significantly outperforms the state-of-the-art LFA models in terms of computational efficiency for predicting the missing data of an HDI matrix with competitive accuracy.

*Index Terms*—High-dimensional and incomplete Data, Latent Factor Analysis, Stochastic Gradient Descent, Fuzzy Proportional Integral Derivation.

## I. INTRODUCTION

Industry applications driven by big data universally involve massive nodes, i.e., users and items in recommender system [14, 19, 20, 50]. With the explosive growth of the involved nodes, it is unable to obtain full interactions among them [1, 3, 6]. For instance, a user only interacts a tiny fraction of items in a recommender system [30, 38, 44]. Consequently, the resultant interactions can be described by a high-dimensional and incomplete (HDI) matrix [10, 13, 17], whose characteristic of high-dimensional corresponding to numerous nodes and high-incomplete corresponding to the interactions among them. Such a matrix only contains a few known entries (the known interactions), yet the remaining vast majority of unobserved entries are unknown instead of zeroes (the unknown interactions) [21-24, 32, 33, 35].

An HDI matrix contains a great deal of serviceable knowledge regarding various patterns like user-item preferences in a recommender system [43-46]. Hence, how to process it effectively for various downstream tasks is an essential issue. In recent years, a lot of valuable researches have emerged [39-42]. Amongst these noteworthy methods, a latent factor analysis (LFA) model has attracted general attention owing to its high efficiency and expandability [36]. Given an HDI matrix, an LFA model trains the desired LFs by embedding the row and column entities of a target HDI matrix into a low-dimensional LF space, and constructs a low rank approximation to the target HDI matrix based on its known data only [8, 29].

Prior studies on existing LFA models [1, 2, 18] demonstrate that a standard stochastic gradient descent (SGD) algorithm is high scalability and efficiency to build an LFA model. However, a standard SGD algorithm commonly encounters the problem of slow convergence because a standard SGD algorithm learns a latent factor relying on the stochastic gradient of current instance error only without considering past update information.

As unveiled by prior work [2, 4, 5], classic control models in automatic control can stabilize a system quickly by utilizing its historical errors. Especially, a proportional-integral-derivative (PID) controller [7] is the most commonly adopted to achieve this goal due to its simplicity and effectiveness. Hence, enlightened by this discovery, a PID controller have been introduced into a machine learning model improve its convergence rate from model-driven perspective.

However, existing studies [47, 48] all adopt a standard PID controller to consider the past update information. As shown in prior researches [48], the control parameters of a standard PID controller need to be manual tuning, thereby perplexing the issue of parameter tuning in different applications. Moreover, they are all always fixed during the control process, which greatly restricts its ability to solve a complex nonlinear problem [12, 25, 31], i.e., implementing LFA on a HDI matrix is a bilinear and non-convex optimization problem. Note that a fuzzy PID controller [9, 11] is able to make the control parameters self-adaptation by adopting the rules of fuzzy reasoning, thereby achieving performance gain on complex nonlinear problems. Additionally, the hyper-parameters like lambda and learning rate in LFA model requires gird search, this issue also affects model performance when processing an industrial HDI matrix. Thereby Li *et al.* [47] propose to multiply the hyper-parameters into the gain parameters expression and to adapt all parameters at the same time. Thus, it is also a critical problem to apply fuzzy method to realized hyper-parameters adaptive. Inspired by above discovery, the following research question are put forward:

✧ J. L. Li is with the School of Computer Science and Technology, Chongqing University of Posts and Telecommunications, Chongqing 400065, China, and also with the Chongqing Key Laboratory of Big Data and Intelligent Computing, Chongqing Engineering Research Center of Big Data Application for Smart Cities, and Chongqing Institute of Green and Intelligent Technology, Chinese Academy of Sciences, Chongqing 400714, China (e-mail: appleli_li@163.com).
✧ Y. Yuan is with the College of Computer and Information Science, Southwest University, Chongqing 400715, China (e-mail: yuanyekl@gmail.com,).

***RQ.*** Is it possible to incorporate past update information into SGD following the principle of a fuzzy PID controller, and achieve all parameters adaptation with fuzzy rules, thereby improving its convergence rate?

To response this question, this paper innovatively proposes a <u>F</u>uzzy <u>P</u>ID-incorporated <u>S</u>GD (FPS) algorithm, which can refine the current instance error by considering the past update information in an efficient way and realize all parameters adaptation following the principle of a fuzzy PID. Accordingly, an FPS-based LFA model is further proposed.

This paper mainly contributes in the following perspectives: a) it presents an FPS model, it incorporates an FPS algorithm for implementing fast LFA on HDI data, and b) Empirical studies on four HDI matrices testify the excellent performance of the proposed FPS model in comparison with state-of-the-art models.

Section II states the preliminaries. Section III presents the methodology. Section IV conducts the empirical studies. Finally, Section V concludes this paper.

## II. PRELIMINARIES

### A. Problem Formulation

As our fundamental data source, an HDI matrix is defined as [26-28]:

***Definition* 1:** Given two large node sets $M$ and $N$ denote the user and item sets respectively, $R \in R^{|M| \times |N|}$'s shows the rating matrix whose each value $r_{m,n}$ is user $m$'s preference on item $n$, $\Lambda$ and $\Gamma$ is the known and unknown entity sets of $R$, $R$ is an HDI matrix if $|\Lambda| \ll |\Gamma|$ [1].

***Definition* 2.** Given $R$ and $\Lambda$, a low-rank estimation $\hat{R}=XY^T$ defined on $R_\Lambda$ represent a standard LFA model, LF matrices $X^{|M| \times |f|}$ and $Y^{|N| \times |f|}$ represent $M$ and $N$, and $f \ll \min\{|M|, |N|\}$ [13].

Thus, the loss function with Euclidean distance is formulated as [18]:

$$\varepsilon(X,Y) \triangleq \left( (R-\hat{R})^2 + \lambda(\|X\|_F^2 + \|Y\|_F^2) \right)$$
$$= \sum_{r_{m,n} \in \Lambda} \left( (r_{m,n} - \hat{r}_{m,n})^2 + \lambda \|x_m\|_2^2 + \lambda \|y_n\|_2^2 \right), \quad (1)$$

where $r_{m,n}$ and $\hat{r}_{m,n}$ denotes the single element in $R$ and $\hat{R}$, $x_m$ and $y_n$ are $m$-th and $n$-th row vector of $X$ and $Y$, $\lambda$ denotes regularization coefficient, and $\|\cdot\|_2$ computes the $L_2$ norm of a vector.

### B. An SGD Algorithm for LFA

According to previous studies, SGD algorithm has been widely applied in optimization problems due to its effectiveness, interpretability, and simplicity. Thus, the implementation of the objective function (1) with SGD algorithm takes the form:

$$\arg\min_{X,Y} \varepsilon \stackrel{SGD}{\Rightarrow} \forall r_{m,n} \in \Lambda: \begin{cases} x_m \leftarrow x_m - \eta \frac{\partial \varepsilon_{m,n}}{\partial x_m}, \\ y_n \leftarrow y_n - \eta \frac{\partial \varepsilon_{m,n}}{\partial y_n}; \end{cases} \quad (2)$$

where $\eta$ is learning rate, $\frac{\partial \varepsilon_{m,n}}{\partial x_m}$ and $\frac{\partial \varepsilon_{m,n}}{\partial y_n}$ is gradient of $x_m$ and $y_n$; and the instant objective $\varepsilon_{m,n}$ is given as:

$$\varepsilon_{m,n} = (r_{m,n} - \langle x_m, y_n \rangle)^2 + \lambda \|x_m\|_2^2 + \lambda \|y_n\|_2^2. \quad (3)$$

Note that learning error $e_{m,n} = r_{m,n} - \langle x_m, y_n \rangle$, then, we folding (2)-(3), our final standard SGD-based LFA model as follow:

$$\arg\min_{X,Y} \varepsilon \stackrel{SGD}{\Rightarrow} \forall r_{m,n} \in \Lambda: \begin{cases} x_m \leftarrow x_m + \eta \cdot (e_{m,n} \cdot y_n - \lambda \cdot x_m), \\ y_n \leftarrow y_n + \eta \cdot (e_{m,n} \cdot x_m - \lambda \cdot y_n). \end{cases} \quad (4)$$

### C. Principle of Fuzzy PID Controller

Fuzzy adaptive control helps complex control systems adapt their parameters, and greatly reduces the complexity of control system regulation. Based on this, fuzzy PID has been put forward [15, 16, 34, 37].

Note that $\iota$ denotes the time point of fuzzy PID controller. Let $e^{(t)}$ represent the current error at the $t$-th point, mathematically, there is:

$$\tilde{e}^{(t)} = K_P e^{(t)} + K_I \sum_{k=0}^{t} e^{(k)} + K_D \left( e^{(t)} - e^{(t-1)} \right), \quad (5)$$

where $e_c = e^{(t)} - e^{(t-1)}$ represents the rate of change of error; $K_P$, $K_I$ and $K_D$ are the correction parameters that need adaptive by fuzzy rules at $t$-th time point; $\tilde{e}^t$ is the output of the fuzzy PID controller at time $\iota$.

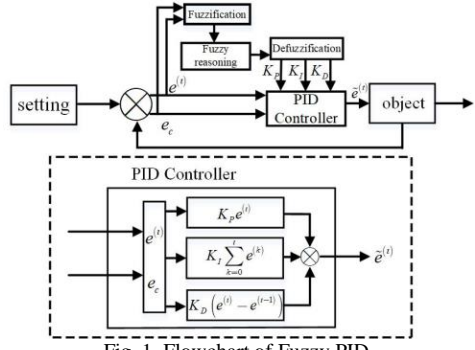

Fig. 1. Flowchart of Fuzzy PID.

III. METHODS

*A. FPS-based Learning Error Refinement*

Fig. 1 indicates that fuzzy-PID uses all errors from beginning to current to revise the learning error of *t*-th time. Thus, we execute a controller for each instance $r_{m,n} \in \Lambda$, and consider each iteration as a time point in controller. We execute $|\Lambda|$ different controllers in LFA model, each controller is responsible for processing an error sequence. Through these operations, we utilize the controller to refine the learning error at the *t*-th time point on each instance as follows:

During the *t*-th training iteration: $\arg\min_{X,Y} \varepsilon \overset{SGD}{\Rightarrow}$

$$\forall r_{m,n} \in \Lambda : \begin{cases} x_m \leftarrow x_m + \eta \cdot \left( e_{m,n}^{(t)} \cdot y_n - \lambda \cdot x_m \right), \\ y_n \leftarrow y_n + \eta \cdot \left( e_{m,n}^{(t)} \cdot x_m - \lambda \cdot y_n \right); \end{cases} \quad (6)$$

where $e_{m,n}^{(t)}$ is the learning error of instance $r_{m,n}$ at *t*-th training iteration, $\tilde{e}_{m,n}^{(t)}$ is the refinement learning error corresponding to $e_{m,n}^{(t)}$. From (5) and (6), we express the fuzzy-PID based error refining rule as follow:

$$\tilde{e}_{m,n}^{(t)} = K_P e_{m,n}^{(t)} + K_I \sum_{k=0}^{t} e_{m,n}^{(k)} + K_D \left( e_{m,n}^{(t)} - e_{m,n}^{(t-1)} \right). \quad (7)$$

Through the above analysis, we obtain the fuzzy-PID incorporated SGD-based learning scheme for LFA model as formula (8):

$$\begin{aligned}
\tilde{e}_{m,n}^{(t)} &= K_P e_{m,n}^{(t)} + K_I \sum_{k=0}^{t} e_{m,n}^{(k)} + K_D \left( e_{m,n}^{(t)} - e_{m,n}^{(t-1)} \right), \\
x_m &\leftarrow x_m + \eta \cdot \left( \tilde{e}_{m,n}^{(t)} \cdot y_n - \lambda \cdot x_m \right), \\
y_n &\leftarrow y_n + \eta \cdot \left( \tilde{e}_{m,n}^{(t)} \cdot x_m - \lambda \cdot y_n \right);
\end{aligned} \quad (8)$$

Formula (8) has two parameters $\eta$ and $\lambda$ need to adjust expect gain parameters $K_P$, $K_I$, and $K_D$. Through the analysis of SGD-LFA model, the dynamic changes of $\eta$ and $\lambda$ with the feedback results of model can help improve the convergence efficiency and reduce the time cost of manual adjustment. Therefore, we combine regularization coefficient $\lambda$ multiply learning rate $\eta$ into a new constant term $\varphi$, and multiply the learning rate $\eta$ into the control expression like Formula (9):

$$\begin{aligned}
x_m &\leftarrow (1 - \eta \cdot \lambda) \cdot x_m + \left( \eta \cdot K_P e_{m,n}^{(t)} + \eta \cdot K_I \sum_{k=0}^{t} e_{m,n}^{(k)} + \eta \cdot K_D \left( e_{m,n}^{(t)} - e_{m,n}^{(t-1)} \right) \right) \cdot x_m \\
&\leftarrow (1 - \varphi) \cdot x_m + \left( \tilde{K}_P e_{m,n}^{(t)} + \tilde{K}_I \sum_{k=0}^{t} e_{m,n}^{(k)} + \tilde{K}_D \left( e_{m,n}^{(t)} - e_{m,n}^{(t-1)} \right) \right) \cdot x_m, \\
y_n &\leftarrow (1 - \eta \cdot \lambda) \cdot y_n + \left( \eta \cdot K_P e_{m,n}^{(t)} + \eta \cdot K_I \sum_{k=0}^{t} e_{m,n}^{(k)} + \eta \cdot K_D \left( e_{m,n}^{(t)} - e_{m,n}^{(t-1)} \right) \right) \cdot y_n \\
&\leftarrow (1 - \varphi) \cdot y_n + \left( \tilde{K}_P e_{m,n}^{(t)} + \tilde{K}_I \sum_{k=0}^{t} e_{m,n}^{(k)} + \tilde{K}_D \left( e_{m,n}^{(t)} - e_{m,n}^{(t-1)} \right) \right) \cdot y_n,
\end{aligned} \quad (9)$$

where $\varphi$ represents product of $\eta$ and $\lambda$; $\tilde{K}_P$, $\tilde{K}_I$, and $\tilde{K}_D$ represent the proportional, integral, and derivative parameters after multiplied by $\eta$.

Hence, there are four parameters here which are $\varphi$, $\tilde{K}_P$, $\tilde{K}_I$, and $\tilde{K}_D$ need to adaptation according fuzzy rules in our model.

*B. Design of Fuzzy Reasoning Process*

Note that the root mean squared error (RMSE) is decreasing until the model converges. Thus, we choose the negative derivative of RMSE rather than learning error as the input of fuzzy control. The formula of RMSE shown as:

$$RMSE^t = \sqrt{\left(\sum_{r_{m,n}\in\Omega} |r_{m,n} - \hat{r}_{m,n}|^2\right)\bigg/|\Omega|};$$

$$A^t = RMSE^{t-1} - RMSE^t, \tag{10}$$

where $\hat{r}_{m,n}$ is the prediction to $r_{m,n}$ generated on test data, $\Omega$ is the validation set, $t$ is the iteration count, $RMSE^{(t)}$ is the RMSE at $t$-th iteration, $A^{(t)}$ is negative derivative of RMSE at $t$-th iteration.

**Fuzzification:** We range $A^{(t)}$ based on empirical values and divide it into several subintervals. Then, we observe which subinterval it belongs to after computing $A^{(t)}$ (e.g. $A^{(t)} \in [A_1, A_2]$, if $A_1 \leq A^{(t)} < A_2$).

Then, the triangular membership function is used to transform $A^{(t)}$ into fuzzy quantities. That is to say, membership function calculated the membership degrees of $A^{(t)}$ belonging to A and B, respectively (i.e. the membership degree is $D_{A1}$ and $D_{A2}$).

**Fuzzing reasoning:** As for the fuzzy reasoning process, fuzzy rules of $\varphi$ should be proportional to the change in learning error; fuzzy rules of $\tilde{K}_P$, $\tilde{K}_I$, and $\tilde{K}_D$ should satisfy the different functions of each item in the control process. From the prior experience, we obtain the fuzzy reasoning table of parameters following the values of $A^{(t)}$ as Table I. In case that interval points $A_1 \sim A_5$, $\varphi_1 \sim \varphi_5$, $P_1 \sim P_5$, $I_1 \sim I_5$, and $D_1 \sim D_5$ are called the membership value and they are positive numbers.

TABLE I. Fuzzy reasoning table and data relationship.

| $A^{(t)}$ | $A_1$ | $A_2$ | $A_3$ | $A_4$ | $A_5$ |
|---|---|---|---|---|---|
| $\varphi$ | $\varphi_1$ | $\varphi_2$ | $\varphi_3$ | $\varphi_4$ | $\varphi_5$ |
| $K_P$ | $P_1$ | $P_2$ | $P_3$ | $P_4$ | $P_5$ |
| $K_I$ | $I_1$ | $I_2$ | $I_3$ | $I_4$ | $I_5$ |
| $K_D$ | $D_1$ | $D_2$ | $D_3$ | $D_4$ | $D_5$ |

($A_1<A_2<A_3<A_4<A_5$; $\varphi_1<\varphi_2<\varphi_3<\varphi_4<\varphi_5$; $P_1<P_2<P_3<P_4<P_5$; $I_1>I_2>I_3>I_4>I_5$; $D_1>D_2>D_3>D_4>D_5$)

**Defuzzification:** As for the fuzzy reasoning analysis, fuzzy rules of $\varphi$, $\tilde{K}_P$, $\tilde{K}_I$, and $\tilde{K}_D$ are presented in Table II (i.e. as $A^{(t)} \in [A_1, A_2]$, then $\varphi \in [\varphi_1, \varphi_2]$, $\tilde{K}_P \in [P_1, P_2]$, $\tilde{K}_I \in [I_1, I_2]$, and $\tilde{K}_D \in [D_1, D_2]$).

According to the above analysis, we have the input membership value, its membership degree, and the input membership value. Finally, the output of the fuzzy controller can be calculated as follow:

$$\varphi = D_{A1}\cdot\varphi_1 + D_{A2}\cdot\varphi_2, \quad K_P = D_{A1}\cdot P_1 + D_{A2}\cdot P_2,$$
$$K_I = D_{A1}\cdot I_1 + D_{A2}\cdot I_2, \quad K_D = D_{A1}\cdot D_1 + D_{A2}\cdot D_2; \tag{11}$$

The above steps have explained the procedure of fuzzification, fuzzy reasoning, and defuzzification of fuzzy PID, and the output can be used as parameters to participate in the $t$-th training as shown in formula (9).

## IV. EXPERIMENTAL RESULTS AND ANALYSIS

### A. General Settings

**Evaluation Protocol**. To measure the accuracy of FPS model, we adopt the root mean squared error (RMSE) as evaluation metrics, the lower RMSE value represent higher prediction accuracy of desired matrices. Thus, the formula of RMSE is shown as below:

$$RMSE = \sqrt{\left(\sum_{r_{m,n}\in\Phi} |r_{m,n} - \hat{r}_{m,n}|^2\right)\bigg/|\Phi|}$$

where $\hat{r}_{m,n}$ is the prediction value to $r_{m,n}$ generated on test data, $\Phi$ is the testing set.

**Datasets.** The detail of four industrial HDI matrices adopted in our experiments are shown as below:

TABLE II. Experimental dataset details.

| No. | Name | Row | Column | Known Entries | Density |
|---|---|---|---|---|---|
| D1 | ML10M [48] | 71,567 | 10,681 | 10,000,54 | 0.31% |
| D2 | ML20M [48] | 138,493 | 26,744 | 20,000,263 | 0.54% |
| D3 | Douban [48] | 129,490 | 58,541 | 16,830,839 | 0.22% |
| D4 | Jester [48] | 124,113 | 150 | 5,865,235 | 31% |

In the experiment process, we divide 70% of all known entry set for each HDI date into a training set, 20% data rest into a testing set, and the rest into a validation set.

a) **General Settings**. More parameter settings for the LFA model and fuzzy reasoning process are shown as follows: a) The target LF matrices $X$ and $Y$ in all LFA models is initialized randomly; b) After considering the accuracy and efficiency of LFA model, the dimension of LF space $f$ is set as 20 uniformly; c) According to the experience value, we selected a reasonable initialization value for the parameters that need to be adaptive by fuzzy rules: $\varphi = 0.00012$, $\tilde{K}_P = 0.005$, $\tilde{K}_I = 0.000001$, $\tilde{K}_P = 0.0002$; and d) Based on the empirical obtained after training, we set membership value in fuzzy PID $A_1 \sim A_5$ is (0.0001, 0.0002, 0.0003, 0.0004, 0.0005), $\varphi_1 \sim \varphi_5$ is (0.00006, 0.00007, 0.00008, 0.00009, 0.0001) $P_1 \sim P_5$ is (0.004, 0.0045, 0.005, 0.0055, 0.006), $I_1 \sim I_5$ is (0.0000009, 0.0000008, 0.0000007, 0.0000006, 0.0000005), $D_1 \sim D_5$ is (0.0000072, 0.0000064, 0.0000056, 0.0000048, 0.000004).

## B. Comparison against State-of-the-art Models

In this section, we give the comparative experimental results of FPS-based LFA model with several state-of-art LFA models. Table III records the details of all compared models. Table IV records the detailed performance of M1-4 on D1-4. From those results, we have following findings:

a) **M1's computational efficiency is surprising and significantly higher than its peers.** For instance, FPS takes 31.6 seconds to achieve the lowest RMSE on D1, which is about 55.74% (i.e., ($Cost_{high}$-$Cost_{low}$)/$Cost_{high}$) lower than M2's 71.4 seconds, 60.35% lower than M3's 79.7 seconds, 76.48% lower than M4's 134.4 seconds.

**M1 also achieves compared results in terms of prediction accuracy for missing data.** For instance, on D1, M1's RMSE is 0.7918, which is 0.13% lower than 0.7929 by M2, 0.37% lower than 0.7948 by M3, 0.26% lower than 0.7939 by M4.

TABLE III. Details of compared models.

| Model | Name | Description |
|---|---|---|
| M1 | FPS | The proposed FPS model of this study. |
| M2 | PSL | A standard PID-incorporated SGD-based LFA model [48]. |
| M3 | PIDoptimizer | An LFA model with a PID-based stochastic optimizer [49]. |
| M4 | SGD-LFA | A standard SGD-based LFA model. |

TABLE IV. Lowest RMSE and their corresponding total time cost (Secs).

| Case | | M1 | M2 | M3 | M4 |
|---|---|---|---|---|---|
| D1 | RMSE: | **0.7918**$_{\pm 5E-4}$ | 0.7929$_{\pm 3E-4}$ | 0.7948$_{\pm 3E-4}$ | 0.7939$_{\pm 5E-5}$ |
| | Time: | **31.6**$_{\pm 1.8}$ | 71.4$_{\pm 1.5}$ | 79.7$_{\pm 3.6}$ | 134.4$_{\pm 1.2}$ |
| D2 | RMSE: | **0.7842**$_{\pm 5E-4}$ | 0.7850$_{\pm 5E-4}$ | 0.7882$_{\pm 1E-4}$ | 0.7866$_{\pm 5E-4}$ |
| | Time: | **71.0**$_{\pm 1.3}$ | 149.4$_{\pm 1.8}$ | 180.9$_{\pm 2.5}$ | 305.2$_{\pm 1.5}$ |
| D3 | RMSE: | **0.7246**$_{\pm 3E-4}$ | 0.7252$_{\pm 5E-4}$ | 0.7290$_{\pm 5E-4}$ | 0.7257$_{\pm 5E-4}$ |
| | Time: | **76.5**$_{\pm 1.6}$ | 150.1$_{\pm 1.6}$ | 167.5$_{\pm 2.0}$ | 336.8$_{\pm 1.3}$ |
| D4 | RMSE: | **1.0054**$_{\pm 4E-4}$ | 1.0090$_{\pm 5E-4}$ | 1.0067$_{\pm 4E-4}$ | 1.0077$_{\pm 4E-4}$ |
| | Time: | **3.1**$_{\pm 1.1}$ | 4.6$_{\pm 0.7}$ | 6.5$_{\pm 2.8}$ | 8.0$_{\pm 0.7}$ |

## V. CONCLUSIONS

In this article, we have proposed an improved fuzzy-PID incorporated SGD LFA model. It contains two steps: 1) utilizing present, history, and tend of feedback to reconstruct error based on PID controller; and 2) creating an appropriate fuzzy table and adapting all parameters with it. In conclusion, FPS-based LFA model has a pretty explanatory effect on the LFA model, and avoids highly expensive tuning process of gain parameters reasonably.